\documentclass[aps,prc,twocolumn,superscriptaddress,showpacs,nofootinbib,floatfix]{revtex4}
\usepackage{graphicx}

\newcommand{\affCENBG}{\affiliation{Centre d'\'Etudes Nucl\'eaires de Bordeaux Gradignan, UMR~5797 CNRS/IN2P3~- Universit\'e de Bordeaux, 19 Chemin du Solarium, CS 10120, F-33175 Gradignan, France}}
\newcommand{\affIFIC}{\affiliation{Instituto de F{\'i}sica Corpuscular, CSIC-Universidad de Valencia, E-46071 Valencia, Spain}}
\newcommand{\affDebrecen}{\affiliation{Institute of Nuclear Research of the Hungarian Academy of Sciences, Debrecen, Hungary}}
\newcommand{\affRIKEN}{\affiliation{RIKEN Nishina Center, 2-1 Hirosawa, Wako, Saitama 351-0198, Japan}}
\newcommand{\affTokio}{\affiliation{Department of Physics, University of Tokyo, 7-3-1 Hongo, Bunkyo-ku, Tokyo 113-0033, Japan}}
\newcommand{\affOsaka}{\affiliation{Department of Physics, Osaka University, Toyonaka, Osaka 560-0043, Japan}}
\newcommand{\affOsakab}{\affiliation{Research Center for Nuclear Physics, Osaka University, Ibaraki, Osaka 567-0047, Japan}}
\newcommand{\affSurrey}{\affiliation{Department of Physics, University of Surrey, Guildford GU2 7XH, Surrey, United Kingdom}}
\newcommand{\affChile}{\affiliation{Comisi{\'o}n Chilena de Energ{\'\i}a Nuclear, Casilla 188-D, Santiago, Chile}}
\newcommand{\affCologne}{\affiliation{Institute of Nuclear Physics, University of Cologne, D-50937 Cologne, Germany}}
\newcommand{\affTUM}{\affiliation{Physik Department E12, Technische Universit\"at M\"unchen, D-85748 Garching, Germany}}
\newcommand{\affLegnaro}{\affiliation{Laboratori Nazionali di Legnaro dell'INFN, I-35020 Legnaro (Padova), Italy}}
\newcommand{\affBucarest}{\affiliation{National Institute for Physics and Nuclear Engineering IFIN-HH, PO BOX MG-6, Bucharest Magurele, Romania}}
\newcommand{\affPadova}{\affiliation{INFN Sezione di Padova and Dipartimento di Fisica, Universit\`a di Padova, I-35131 Padova, Italy}}
\newcommand{\affIstanbul}{\affiliation{Department of Physics, Istanbul University, Istanbul 34134, Turkey}}
\newcommand{\affTokyo}{\affiliation{Department of Physics, Tokyo University of Science, Noda, Chiba 278-8510, Japan}}
\newcommand{\affGANIL}{\affiliation{Grand Acc\'el\'erateur National d'Ions Lourds, B.P. 55027, F-14076 Caen Cedex 05, France}}
\newcommand{\affTennessee}{\affiliation{Department of Physics and Astronomy, University of Tennessee, 401 Nielsen Physics Building, 1408 Circle Drive, Knoxville, TN 37996-1200, USA}}
\begin{document}
\renewcommand{\textfraction}{.01}
\headsep 1.7cm

\title{New neutron-deficient isotopes from $^{78}$Kr fragmentation}

\author{B. Blank} \affCENBG
\author{T. Goigoux} \affCENBG
\author{P. Ascher} \affCENBG
\author{M. Gerbaux} \affCENBG
\author{J. Giovinazzo} \affCENBG
\author{S. Gr\'evy} \affCENBG
\author{T. Kurtukian Nieto} \affCENBG
\author{C.~Magron} \affCENBG 
\author{J. Agramunt} \affIFIC
\author{A. Algora} \affIFIC\affDebrecen
\author{V. Guadilla} \affIFIC
\author{A. Montaner-Piza} \affIFIC
\author{A. I. Morales} \affIFIC
\author{S.E.A. Orrigo} \affIFIC
\author{B. Rubio} \affIFIC
\author{D.S.~Ahn} \affRIKEN    
\author{P. Doornenbal} \affRIKEN
\author{N. Fukuda} \affRIKEN
\author{N. Inabe} \affRIKEN
\author{G. Kiss} \affRIKEN
\author{T. Kubo} \affRIKEN
\author{S. Kubono} \affRIKEN
\author{S. Nishimura} \affRIKEN
\author{V. H. Phong} \affRIKEN
\author{H. Sakurai} \affRIKEN\affTokio
\author{Y.~Shimizu} \affRIKEN
\author{P.-A. S\"oderstr\"om} \affRIKEN
\author{T. Sumikama} \affRIKEN
\author{H. Suzuki} \affRIKEN
\author{H. Takeda} \affRIKEN
\author{J. Wu} \affRIKEN
\author{Y.~Fujita} \affOsaka\affOsakab
\author{M. Tanaka} \affOsaka
\author{W. Gelletly} \affIFIC\affSurrey
\author{P. Aguilera} \affChile
\author{F. Molina} \affChile
\author{F. Diel} \affCologne
\author{D. Lubos} \affTUM
\author{G. de Angelis} \affLegnaro
\author{D.~Napoli} \affLegnaro
\author{C. Borcea} \affBucarest
\author{A. Boso} \affPadova
\author{R.B. Cakirli} \affIstanbul
\author{E. Ganioglu} \affIstanbul
\author{J. Chiba} \affTokyo
\author{D. Nishimura} \affTokyo
\author{H. Oikawa} \affTokyo
\author{Y. Takei} \affTokyo
\author{S. Yagi} \affTokyo
\author{K. Wimmer} \affTokio
\author{G. de France} \affGANIL
\author{S. Go} \affTennessee

\begin{abstract}

In an experiment with the BigRIPS separator at the RIKEN Nishina Center, the fragmentation of a $^{78}$Kr beam 
allowed the observation of new neutron-deficient isotopes at the proton drip-line.
Clean identification spectra could be produced and $^{63}$Se, $^{67}$Kr, and $^{68}$Kr
were identified for the first time. In addition, $^{59}$Ge was also observed. Three of these isotopes, 
$^{59}$Ge, $^{63}$Se, and $^{67}$Kr, are potential candidates for ground-state two-proton radioactivity.
In addition, the isotopes $^{58}$Ge, $^{62}$Se, and $^{66}$Kr were also sought but without success. 
The present experiment also allowed the determination of production cross sections for some of the most exotic isotopes.
These measurements confirm the trend already observed that the empirical parameterization of fragmentation
cross sections, EPAX, significantly overestimates experimental cross sections in this mass region.

\end{abstract}


\maketitle

\section{Introduction}

Experiments on nuclei at the limits of nuclear stability allow the testing of the nuclear models under extreme conditions. Thus concepts well
established close to the line of nuclear stability can be checked for their ability to describe nuclear structure over the
whole chart of isotopes. Very often it is found that these concepts have to be modified and new concepts developed.

When moving close to the limits of stability, the mass difference between neighboring nuclei increases.
Therefore new decay channels open and exotic radioactivities appear. For example, close to stability $\beta-\gamma$ decay
is the most common decay mode for radioactive nuclei. When approaching the limits of stability the available energy for
the decay, the $Q$ value, increases and the particle emission threshold, e.g. the proton separation energy, decreases and
$\beta$-delayed particle emission becomes a more and more important decay channel. Close to the proton drip-line, $\beta$-delayed
one-, two-, and three-proton emission have all been observed~\cite{blank08review,pomorski11a,audirac12}.

After crossing the proton drip-line (proton separation energy $S_p < 0$), the direct emission of protons becomes the dominant decay 
channel. While one-proton (1p) radioactivity is today an established decay channel used in many studies of nuclear structure beyond 
the limits of nuclear stability~\cite{blank08review}, two-proton (2p) radioactivity is the most recently discovered radioactivity and 
only a few cases are known at present~\cite{blank08review2p}. Two-proton radioactivity with half-lives of the order of milli-seconds 
was first observed in the region of iron-nickel-zinc with the 2p emitters $^{45}$Fe, $^{48}$Ni, and $^{54}$Zn~\cite{blank08review2p}. 
Just above this region, $^{59}$Ge, $^{63}$Se, and $^{67}$Kr were predicted to be possible new 2p emitters (see e.g.~\cite{brown02a}).

Additional interest in studying proton-rich nuclei arises from the fact that these nuclei often lie on the path of astrophysical
nucleosynthesis processes~\cite{schatz98}. The $rp$-process, a sequence of proton captures and $\beta$ decays, produces many of 
these proton-rich nuclei close to the proton drip-line and a precise knowledge of the properties of these nuclei, i.e. their masses, 
half-lives, and decay properties, is essential for the correct modeling of these processes.

In the region of interest of the present experiment between zinc and krypton ($Z=30-36$), the limits of stability of all odd-$Z$ 
(proton number) elements are believed to be known~\cite{blank95kr,pfaff96}. However, due to the pairing energy, the even-$Z$ elements 
can bind isotopes with even fewer neutrons and thus the two-proton drip-line lies even further away from the valley of stability 
than the one-proton drip-line for odd-$Z$ elements. The drip-line for zinc has clearly been crossed with the observation of
two-proton radioactivity for $^{54}$Zn~\cite{blank05zn54,ascher11}. In an experiment at the NSCL of Michigan State 
University~\cite{stolz05}, $^{60}$Ge and $^{64}$Se were observed for the first time with three and four events, respectively.
The last new isotope in this region, reported after the completion of the present experiment,
is $^{59}$Ge, which was also produced at the NSCL with four counts~\cite{ciemny15}. Finally for krypton, the last known isotope
prior to the present experimental effort was $^{69}$Kr observed at the LISE3 facility of GANIL~\cite{blank95kr}.

In an experiment at the BigRIPS separator of the RIKEN Nishina Center, we produced many proton-rich nuclei with unprecedented 
intensity, some of them for the first time. In the following, we will describe the experimental details, give the results concerning the
production of new isotopes and their production cross sections, and put these results into a general context.

\section{Experimental details}

In a recent experiment at the Radioactive Ion Beam Factory (RIBF) of the RIKEN Nishina Center, the Big\-RIPS 
separator~\cite{bigrips03,bigrips12} was used to fragment a primary $^{78}$Kr beam  at 345 MeV/A with an 
intensity of up to 250~pnA. The primary beam impinged on $^9$Be targets with thicknesses of 5~mm (settings on 
$^{51}$Ni, $^{65}$Br, and $^{64}$Se) and 7~mm (setting on $^{62}$Se). The fragments produced 
in these targets were separated according to their magnetic rigidity in the first dipole of BigRIPS
and their energy loss in a first degrader (focal plane F1, aluminium, 2~mm), before being analyzed again according to their 
magnetic rigidity by the second magnet of Big\-RIPS. A second achromatic degrader (focal plane F5, aluminium, 2~mm) allowed 
us to enhance the selectivity of Big\-RIPS. The second half of BigRIPS was used to measure the time-of-flight of the isotopes transmitted, 
their positions at different focal planes, and their energy loss.  For this purpose, a series of parallel-plate avalanche counters, 
plastic scintillators and multi-sampling ionization chambers was used~\cite{bigrips03,bigrips12}. The BigRIPS separator 
allowed us to separate the isotopes of interest from the bulk of less exotic nuclei produced and to detect and identify the 
most exotic species by means of the $\Delta$E-ToF-B$\rho$ method. 

However, scattered beam particles, double hits occuring mainly in detectors at the beginning of BigRIPS, incomplete events 
and other problems may yield erroneous identification parameters. Therefore, in the off-line analysis cuts have been placed 
on signals from various beam-line detectors to remove these events (see~\cite{fukuda13} for details). After these cuts, 
clean identification spectra could be produced. 

The nuclei thus selected were transmitted to the Zero-degree spectrometer at the end of which a set-up for decay spectroscopy
was installed comprising the double-sided silicon strip detector set-up WAS3ABi~\cite{wasabi} for implantation and detection 
of charged particles emitted during the decay of implanted isotopes and the EURICA germanium detector array~\cite{eurica}.
Measurements performed with this set-up will be reported elsewhere.

\section{Results and discussion}

In the present analysis, we used data from four different settings of the BigRIPS separator: (i) a setting optimised on $^{51}$Ni 
to produce well-known $\beta$-delayed proton emitters for detector calibration. The run at this setting lasted for 12~h with an effective data taking 
time of 10.5~h. (ii) a setting on the unbound isotope $^{65}$Br, which lies between two of the isotopes sought namely $^{63}$Se 
and $^{67}$Kr. The runs at this setting lasted about 156~h with an effective counting time of 115~h. (iii) a run at a setting on 
$^{64}$Se for 52~h with an effective data taking time of about 50~h. (iv) a run on a setting on $^{62}$Se 
to search for $^{58}$Ge and $^{62}$Se for 52~h with an effective data taking time of 48~h. 

\begin{figure*}
\begin{center}
\resizebox{.49\textwidth}{!}{\includegraphics{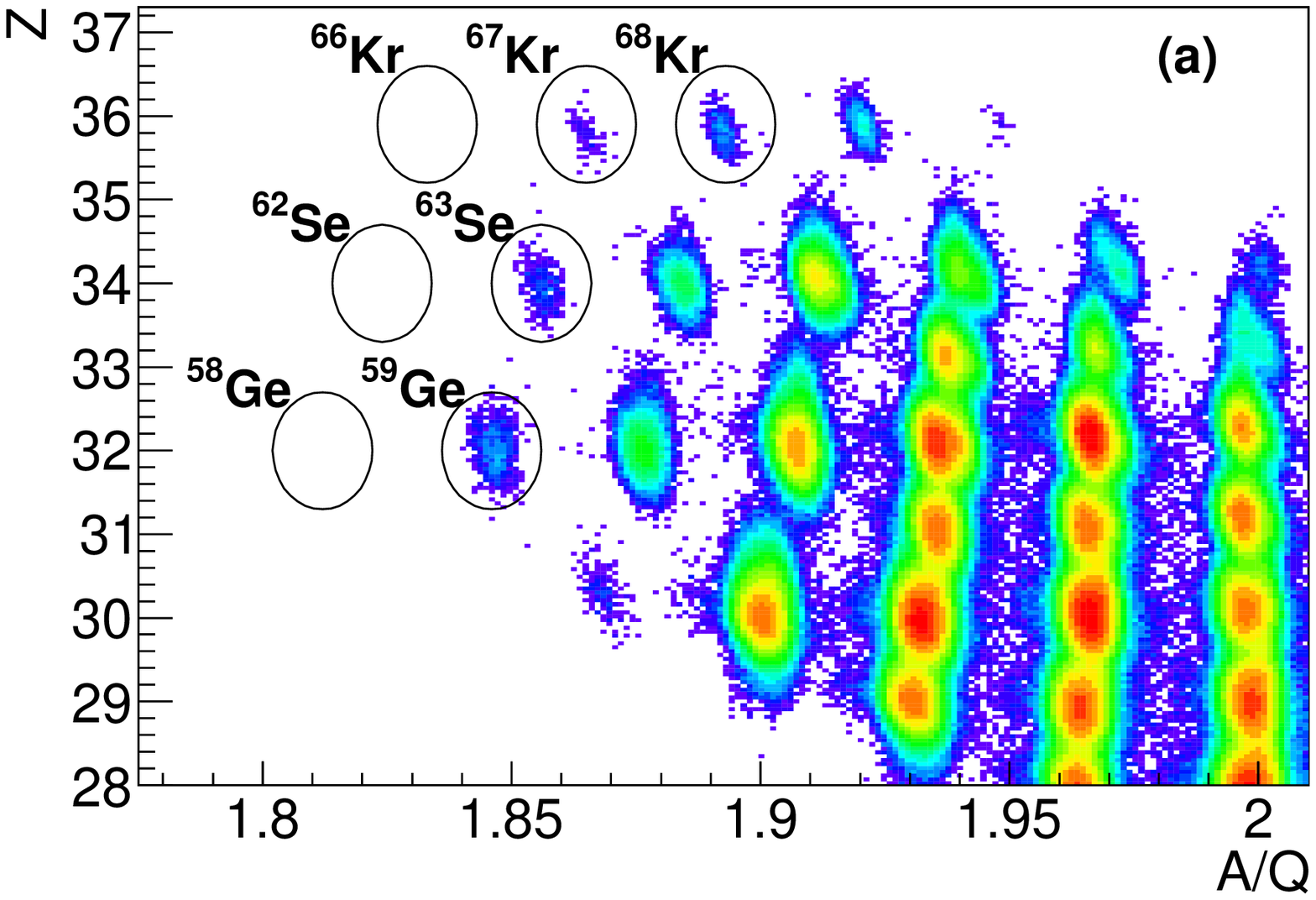}} \ \hfill \
\resizebox{.49\textwidth}{!}{\includegraphics{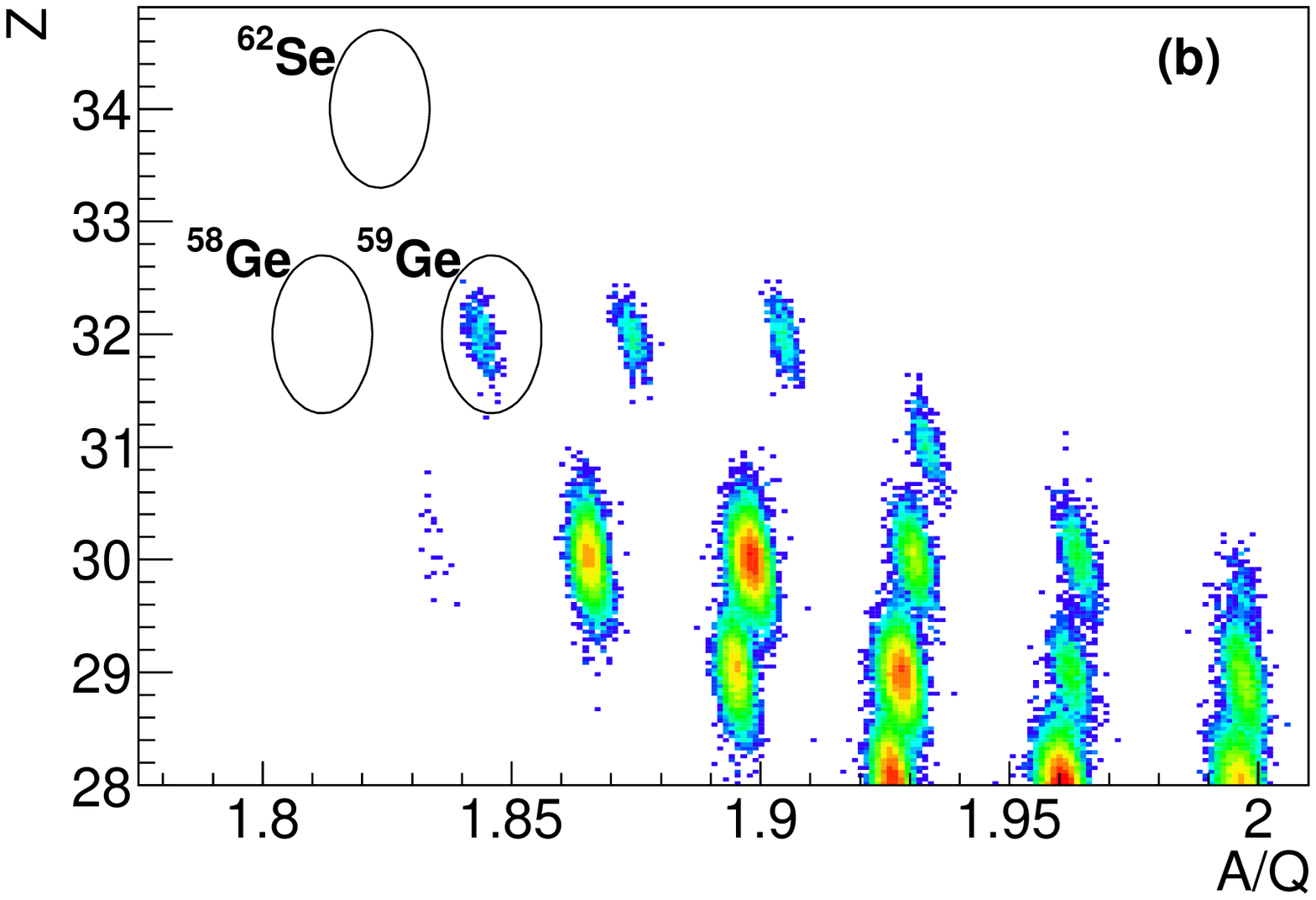}} 
\caption[]{Identification plot of the charge $Z$ of the nuclides as a function of their ratio A-over-Q. (a) Isotopes produced in the
           setting optimized on $^{65}$Br. The isotopes of interest
           are highlighted by a surrounding ellipse and indicated with their names.  $^{63}$Se, $^{67}$Kr, and $^{68}$Kr are
           observed for the first time in the present work. (b) Identification plot for the setting on $^{62}$Se with the isotopes 
           of interest indicated. The absence of $^{58}$Ge, $^{62}$Se, and $^{66}$Kr is evident in (a), whereas the results from the 
           setting presented in (b) allow one to conclude that $^{58}$Ge and $^{62}$Se are unbound and therefore unobserved
           (see text for details).
           
	   }
\label{fig:id}
\end{center}
\end{figure*}

Fig.~\ref{fig:id}(a) presents the identification plot of nuclides at the exit of BigRIPS at its focal plane F7 for the second 
setting, while part (b) is from the last setting. The new isotopes $^{63}$Se, $^{67}$Kr and $^{68}$Kr are indicated. 
The total number of counts for these isotopes, summed over all settings, were 348, 82, and 479 counts, respectively. 
In addition, we produced 1221 nuclei of $^{59}$Ge in the various settings, a factor of 300 more than in a recent 
MSU experiment~\cite{ciemny15}. These numbers are the sums from different settings with greatly different transmissions 
for the different isotopes. Therefore, these numbers should not be compared directly.

The latest Atomic Mass Evaluation (AME2012)~\cite{ame2012} predicts $^{59}$Ge to be two-proton unbound with a two-proton 
separation energy ($S_{2p}$) of -1.66(36)~MeV. Brown {\it et al.}~\cite{brown02a} give a value of -1.16(14)~MeV. We use a simple di-proton 
barrier penetration model which usually yields half-lives that are too short, because it does not take into account correlations 
between the two protons and a proton-proton resonance energy. However, it gives a first guess for barrier penetration half-lives.
Using the first value yields a barrier-penetration half-life of 10~$\mu$s with upper and lower limits of 39~ms 
and 26~ns, respectively, a rather large half-life range which includes the flight time through the BigRIPS separator
of about 410~ns. The separation energy of Brown {\it et al.} gives a half-life range between 0.04~s and 420~s with a 
central value of 2.6~s. Such a long half-life for barrier penetration would clearly indicate that this isotope decays
by $\beta$ decay rather than by 2p emission.

In the case of $^{63}$Se, AME2012 does not give any mass value and thus no separation energy. Brown {\it et al.}
predict a value of -1.51(14)~MeV yielding a half-life range between 0.3~ms and 0.24~s with 6.9~ms for the central
value of the separation energy, which could be compatible with 2p radioactivity.

Finally, for $^{67}$Kr AME2012 again does not give any value, whereas Brown {\it et al.} predict $S_{2p}$~=
\mbox{-1.76(14)~MeV}. This value allows us to determine a barrier-penetration half-life between 62~$\mu$s and 17~ms with a 
central value of 0.83~ms. Evidently, this nucleus seems to be the most promising candidate for 2p radioactivity.
$^{68}$Kr is predicted to have an $S_{2p}$ value of -0.62(14)~MeV~\cite{brown02a}, too small for a possible 2p emission.

The rates observed in the present experiment make it possible to study the decay of these nuclei. In particular, it will be
interesting to see whether $^{59}$Ge, $^{63}$Se, and $^{67}$Kr are indeed 2p emitters. The analysis of the decay of these 
isotopes as well as many others produced in the present experiment with high statistics is ongoing and the results will be 
published elsewhere. 

\begin{figure}[tth]
\begin{center}
\resizebox{.5\textwidth}{!}{\includegraphics[angle=-90]{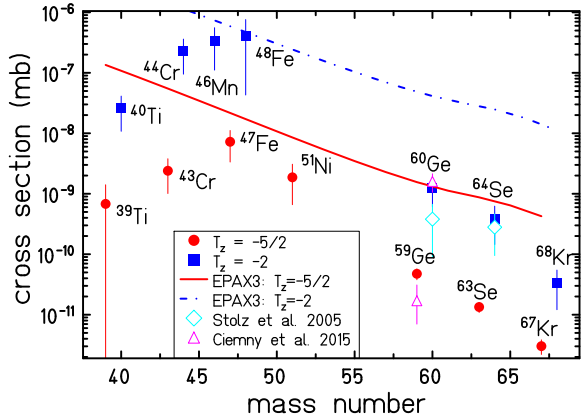}}
\end{center}
\caption[]{Comparison of experimental cross sections and predictions from the EPAX3 parameterization~\cite{epax3} for $T_z$~= 
           -5/2 and -2 nuclei. Details of the data selected for this figure are given in the text and in Table~I.
	       Stolz {\it et al.}, 2005:~\cite{stolz05}, Ciemny {\it et al.}, 2015:~\cite{ciemny15}.}
\label{fig:cross}
\end{figure}

The present experiment allowed us to determine the production cross sections for nuclei transmitted close to the central beam line. 
We limit the data to these nuclei because they have calculated ion optical transmissions between 20\% for the lighter, less exotic nuclei 
(see Fig.~\ref{fig:cross}) and 90\% for fragments closer to the projectile. These relatively high transmission values ensure that
the uncertainties due to the momentum distribution of the fragments and the separator transmission are minimal. For the determination 
of the fragmentation cross sections we used only the setting for $^{65}$Br and the calibration run optimized for $^{51}$Ni. For the 
settings on $^{62}$Se and $^{64}$Se, the slits of BigRIPS were significantly narrower thus introducing much larger errors
for the isotope transmission.  

The cross sections were determined by means of the number of nuclides detected $N_{iso}$, the primary beam intensity $N_{beam}$, 
the transmission through BigRIPS of the different isotopes $T_{iso}$, and the target thickness $d_{tar}$ according to the following formula:
$$
\sigma = \frac{N_{iso}}{N_{beam}}\frac{A}{d_{tar} N_A}\frac{1}{T_{iso}}
$$
with $N_A$ being Avogadro's constant and $A$ the molar weight of the target. The isotopes of interest were counted with
the BigRIPS standard detection set-up (statistical errors only), while the beam intensity was measured with scattered beam
particles by means of scintillation detectors in the vicinity of the production target. This measurement was calibrated at two 
instances during the beam time. The difference between these two calibrations (of the order of 10\%) was used 
as the uncertainty in the beam intensity. The isotopes produced in the target have a certain probability to be destroyed
in a secondary reaction in the production target itself or in the degraders (we neglect here possible interactions in the
detectors the thicknesses of which are much smaller than for the targets and the two degraders). Although the total reaction 
probabilities determined with the formula of Kox {\it et al.}~\cite{kox87} vary a little from one nucleus to the other, 
we use an average secondary reaction probability of 14\% for all nuclei and correct the counting rates for these secondary reactions.
We consider that this correction, already small compared to other corrections, has a negligible uncertainty.
A dead-time of the order of 8\% deduced from the ratio of the number of BigRIPS triggers and those accepted by the data acquisition 
and determined on a run-by-run basis was also taken into account. 

Transmissions were calculated with the LISE++ simulation tool~\cite{lise_code} and the LIESCHEN code~\cite{lieschen} developed
in the framework of the simulation of the FRS~\cite{schmidt87} at GSI. The two codes differ in their degree of sophistication
with the LISE++ code being more elaborate with e.g. the momentum distributions including an exponential tail~\cite{suzuki13}.
However, both codes give results which differ on average by only 12\% for the nuclei in Fig.~\ref{fig:cross}, with the LIESCHEN 
code yielding somewhat lower transmission values. The largest discrepancy in the transmissions from the two codes was 25\%. 
For the cross section calculations, we used the results from the more sophisticated LISE++ code. For the uncertainty, we used the 
difference of the two codes and a smooth additional error that takes into account the general fact that the transmissions are more 
uncertain if they are small. To do so, we employed a smooth function yielding a relative uncertainty of 10\% for a transmission of 90\%
and reaching 40\% of uncertainty for a transmission of 20\%. This is a somewhat arbitrary function, but we believe
that it reflects correctly the fact that the uncertainties increase significantly with the decrease of the transmission. 
The target thickness was measured by means of the $B\rho$ of the primary beam with and without the targets. The 
$B\rho$ measurement precision is of the order of 10$^{-3}$ which leads to an uncertainty in the target thickness of 50~$\mu$m.

Figure~\ref{fig:cross} presents these cross sections and compares them with the predictions of EPAX3, an empirical parameterization 
of fragmentation cross sections~\cite{epax3}. The numerical values are given in Table~I.
Clearly, EPAX3 overestimates the experimental cross sections by a large factor. 

\begin{table}[htt]
\begin{center}
\caption{Experimental fragmentation cross sections are compared to EPAX3~\cite{epax3}. For the data presented in this table, 
         the cross sections are based on ion optical transmissions between 20\% and 90\% for the settings used to produce these nuclei.
         The data for $^{39}$Ti to $^{51}$Ni
         stem from the setting on $^{51}$Ni, whereas the other data are from the setting on $^{65}$Br. 
         }
\begin{tabular}{lrclrcl}
\hline
\hline
nucleus & ~~exp. & cross & section & ~~EPAX3 & cross & section \\
        &   & (mb) & & & (mb) & \\
$^{39}$Ti  & 6.8$^{+7.3}_{-6.8}$& $\times$ & 10$^{-10}$ & 1.3 & $\times$ & 10$^{-7}$   \\
$^{40}$Ti  & 2.6$^{+1.5}_{-1.5}$& $\times$ & 10$^{-8}$  & 2.7 & $\times$ & 10$^{-6}$   \\
$^{43}$Cr  & 2.4$^{+1.4}_{-1.4}$& $\times$ & 10$^{-9}$  & 5.5 & $\times$ & 10$^{-8}$   \\
$^{44}$Cr  & 2.3$^{+1.3}_{-1.3}$& $\times$ & 10$^{-7}$  & 1.1 & $\times$ & 10$^{-6}$   \\
$^{46}$Mn  & 3.3$^{+2.2}_{-2.2}$& $\times$ & 10$^{-7}$  & 7.4 & $\times$ & 10$^{-7}$   \\
$^{47}$Fe  & 7.2$^{+3.9}_{-3.9}$& $\times$ & 10$^{-9}$  & 2.2 & $\times$ & 10$^{-8}$   \\
$^{48}$Fe  & 4.0$^{+3.6}_{-3.6}$& $\times$ & 10$^{-7}$  & 4.7 & $\times$ & 10$^{-7}$   \\
$^{51}$Ni  & 1.9$^{+1.2}_{-1.2}$& $\times$ & 10$^{-9}$  & 8.5 & $\times$ & 10$^{-9}$   \\
$^{59}$Ge  & 4.8$^{+1.0}_{-1.0}$& $\times$ & 10$^{-11}$ & 1.6 & $\times$ & 10$^{-9}$   \\
$^{60}$Ge  & 1.3$^{+0.7}_{-0.7}$& $\times$ & 10$^{-9}$  & 4.1 & $\times$ & 10$^{-8}$   \\
$^{63}$Se  & 1.3$^{+0.3}_{-0.3}$& $\times$ & 10$^{-11}$ & 8.7 & $\times$ & 10$^{-10}$   \\
$^{64}$Se  & 3.8$^{+2.4}_{-2.4}$& $\times$ & 10$^{-10}$ & 2.4 & $\times$ & 10$^{-8}$   \\
$^{67}$Kr  & 3.0$^{+0.8}_{-0.8}$& $\times$ & 10$^{-12}$ & 4.3 & $\times$ & 10$^{-10}$  \\
$^{68}$Kr  & 3.3$^{+2.1}_{-2.1}$& $\times$ & 10$^{-11}$ & 1.1 & $\times$ & 10$^{-8}$  \\
\hline
\hline
\end{tabular}
\end{center}
\label{tab:cross}
\end{table}

The present experimental cross section for $^{59}$Ge is about a factor of 3 larger than the value determined in Ref.~\cite{ciemny15}
of 1.7$^{+1.4}_{-1.0} \times$ 10$^{-11}$~mb. We believe that these differences are most likely due to the difficulty in correctly
calculating the transmission with models. In order to get reliable transmissions, much more settings have to be run to scan the full
transmissions experimentally. 

We believe that the cross sections determined for the lightest nuclei (in particular $^{39}$Ti to $^{44}$Cr) 
should be treated with some caution. An important input for the transmission calculations with the simulation codes is 
the momentum distribution of the fragments and it is not clear whether the momentum parametrisation used in these codes~\cite{suzuki13} 
is also valid for isotopes this far away from the projectile. If the momentum distributions for these isotopes were larger
than expected, the transmissions would be reduced and larger cross sections would be determined. So, maybe it is timely to 
measure the cross sections for $^{78}$Kr fragmentation in a specially designed experiment where the full momentum 
distributions of the nuclides of interest are scanned.
 
The finding that EPAX in its different versions~\cite{epax1,epax2,epax3}, which differ slightly for neutron-deficient nuclei,
overestimates production cross sections in this region of the chart of nuclei is not new (see e.g. \cite{ciemny15,blank07znge,stolz05}). 
However, it is indeed astonishing that EPAX reproduces experimental cross sections for $^{92}$Mo~\cite{fernandez05} and 
$^{58}$Ni~\cite{blank94ni} fragmentation, while in between these two nuclei, EPAX does not work for $^{70}$Ge~\cite{blank07znge} 
and $^{78}$Kr fragmentation (\cite{ciemny15,stolz05} and present work). A possible explanation could be that this is 
linked to nuclear structure effects with $^{92}$Mo lying on the $N$=50 shell closure and $^{58}$Ni having $Z$=28, whereas the other 
two lie between the $N$, $Z$=28 and $N$, $Z$=50 shells. However, according to our knowledge such an effect has not yet been observed elsewhere.

The search for $^{58}$Ge, $^{62}$Se, and $^{66}$Kr in the settings on $^{65}$Br and $^{62}$Se was unsuccessful. No event 
could be identified for any of these nuclei (see Fig.~\ref{fig:id}). From the numbers of counts for the neighbouring nuclei observed 
in the two settings, the transmissions calculated with the two codes LISE++~\cite{lise_code} and LIESCHEN~\cite{lieschen} 
and trends for production cross sections (on average there is a loss of a factor of 22(9) going from one even-$N$ isotope 
to its more exotic odd-$N$ neighbour as determined from our experimental data for even-$Z$ elements and a loss factor of 3.7(17)
between two even-$Z$ isotopes with the same isospin projection, e.g. between $^{59}$Ge and $^{63}$Se), we determine the following
expected numbers of counts: (i) 1.2(12), 1.6(10), and 0.6(3), for $^{58}$Ge, $^{62}$Se, and $^{66}$Kr using the observed numbers 
of counts for the neighbouring nuclei with one more neutron in the setting for $^{65}$Br; 
(ii) 4.8(30) and 20(12) for $^{62}$Se and $^{58}$Ge in the setting for $^{62}$Se
from the number of counts observed for $^{59}$Ge. These numbers are based on the assumption that the isotopes live longer
than the flight time through the separator and 
also include a loss factor of 2 due to the fact that the even-$N$ isotopes of even-$Z$ elements have roughly a factor 
of 2 lower production cross sections than their odd-$N$ neighbours~\cite{blank94ni,blank07znge}.

These low numbers do not allow us to draw definite conclusions about the lifetimes of these nuclei from their non-observation, 
except most likely for $^{58}$Ge where a life-time limit of about 100~ns can be deduced from the flight-time through the BigRIPS 
separator. However, even for the other nuclei, it is unlikely that they have lifetimes comparable to or longer than the 
flight time through the BigRIPS separator. 

This is corroborated by model predictions. Brown {\it et al.}~\cite{brown02a} predict 2p separation energies of 
-2.38(14)~MeV and -2.76(14)~MeV for $^{58}$Ge and $^{62}$Se, respectively. Unfortunately, no value is given for $^{66}$Kr.
With the simple 2p barrier penetration model used above, we obtain half-lives between 0.07~ns and 1.5~ns for $^{58}$Ge 
with a central value of 0.3~ns and between 0.02~ns and 0.3~ns for $^{62}$Se with a value of 0.07~ns for the central 
value of the separation energy. For the case of $^{66}$Kr, we used the Garvey-Kelson method~\cite{kelson66} to determine 
the 2p separation energy and found $S_{2p}$~= -3.0~MeV yielding a barrier penetration half-life of 0.05~ns.

All these values are more than two orders-of-magnitude shorter than the flight time through the BigRIPS separator of 
typically 410~ns for the most exotic nuclei. Thus the non-observation of $^{58}$Ge, $^{62}$Se, and $^{66}$Kr is 
in line with expectations from mass predictions.

Under the assumption of one count observed, the calculated transmissions and their error bars, as well as the beam 
intensity and its uncertainty in the two settings, we can determine one-sigma cross-section limits of 
1.8$\times$10$^{-13}$~mb for $^{66}$Kr, 8.1$\times$10$^{-14}$~mb for $^{62}$Se, and 7.6$\times$10$^{-14}$~mb for $^{58}$Ge. 

\section{Conclusions}

In the present work, we fragmented a high-intensity primary $^{78}$Kr beam and separated the fragments of interest with
the BigRIPS separator. The experiment allowed us to discover the new isotopes $^{63}$Se, $^{67}$Kr, and $^{68}$Kr.
In addition, $^{59}$Ge was observed with high rates. $^{59}$Ge, $^{63}$Se, and $^{67}$Kr are predicted to have
two-proton separation energies of the order of -1~MeV to -2~MeV and are thus potential two-proton emitters, with 
$^{67}$Kr being the best candidate. As these nuclei are 2p unbound by a significant separation energy, they are 
beyond the drip-line and are expected to be the most neutron-deficient isotopes of their elements that can be observed. 
This appears to be confirmed by the non-observation of $^{58}$Ge, $^{62}$Se, and $^{66}$Kr.

The fragmentation cross sections determined for the observed nuclei confirm again that EPAX overpredicts the 
cross sections at the limits of stability in this region. This clearly calls for an experiment designed to 
measure cross sections in this region systematically in order to try to understand this discrepancy.

\section*{Acknowledgment}

This experiment was carried out under programs NP0702-RIBF4R1-01, NP1112-RIBF82, and NP1112-RIBF93 at the RIBF
operated by RIKEN Nishina Center, RIKEN and CNS, University of Tokyo.
We would like to thank the whole RIBF accelerator staff for the support during 
the experiment and for maintaining excellent beam conditions with record intensities.
This work was supported by the French-Japanese "Laboratoire international associ\'e" FJ-NSP, 
by JSPS KAKENHI Grant Number 25247045, 
by the Istanbul University Scientific Project Unit under project number BYP-53195,
by the UK Science and Technology Facilities Council (STFC) Grant no. ST/F012012/1, 
by the Japanese Society for the Promotion of Science under Grant No. 26 04808,
by MEXT Japan under grant number 15K05104, and
by the Spanish Ministerio de Economia y Competitividad under grants FPA2011-24553 and FPA2014-52823-C2-1-P, 
Centro de Excelencia Severo Ochoa del IFIC SEV-2014-0398; $Junta~para~la~Ampliaci\acute{o}n~de~Estudios$ Programme (CSIC JAE-Doc contract) co-financed by FSE.

\end{document}